\documentclass[prd,aps,twocolumn,nofootinbib,preprintnumbers,superscriptaddress,preprintnumbers,balancelastpage,longbibliography]{revtex4-2}
\usepackage[english]{babel}

\usepackage{amsmath,mathtools,physics,xfrac}
\usepackage{graphicx}
\usepackage{afterpage}
\usepackage{float}
\usepackage{subfigure}
\usepackage{rotating}
\usepackage{multirow}
\usepackage{tabularx}
\usepackage{booktabs}
\usepackage{fancyhdr}
\usepackage[utf8]{inputenc}
\usepackage{theorem}
\usepackage{moreverb}
\usepackage{euscript}
\usepackage{psfrag}
\usepackage{slashed}
\usepackage{mathtools}
\usepackage{makecell}
\usepackage{adjustbox}
\usepackage{dcolumn}
\usepackage{bm}
\usepackage[dvipsnames]{xcolor}
\usepackage{graphics}
\usepackage{hyperref}
\usepackage{chngcntr}
\usepackage{aas_macros}
\usepackage{color,xcolor}
\usepackage[normalem]{ulem}
\hypersetup{
     colorlinks   = true,
     citecolor    = Aquamarine,
     urlcolor     = Aquamarine,
     linkcolor    = Aquamarine
}

\definecolor{darkblue}{rgb}{0.0,0.0,0.75}
\definecolor{darkred}{rgb}{0.6,0.0,0}
\definecolor{darkgreen}{rgb}{0.0,0.6,0.}

\counterwithout{equation}{section}

\usepackage{tikz,xcolor,hyperref}

\definecolor{lime}{HTML}{A6CE39}
\DeclareRobustCommand{\orcidicon}{\hspace{-1mm}
	\begin{tikzpicture}
		\draw[lime, fill=lime] (0,0) 
		circle [radius=0.16] 
		node[white] {{\fontfamily{qag}\selectfont \tiny \,ID}};
		\draw[white, fill=white] (-0.0525,0.095) 
		circle [radius=0.007];
	\end{tikzpicture}
	\hspace{-3mm}
}

\foreach \x in {A, ..., Z}{\expandafter\xdef\csname orcid\x\endcsname{\noexpand\href{https://orcid.org/\csname orcidauthor\x\endcsname}
		{\noexpand\orcidicon}}
}

\keywords{}

\begin{document}

\title{\boldmath Hunting Primordial Black Hole Dark Matter in the Lyman-$\alpha$ Forest}

\author{Akash Kumar Saha\orcidA{}}
\email{akashks@iisc.ac.in}

\affiliation{Centre for High Energy Physics, Indian Institute of Science, C.\,V.\,Raman Avenue, Bengaluru 560012, India}

\author{Abhijeet Singh\orcidB{}}
\email{abhijeets@iisc.ac.in}
\affiliation{Centre for High Energy Physics, Indian Institute of Science, C.\,V.\,Raman Avenue, Bengaluru 560012, India}

\author{Priyank Parashari\orcidC{}}
\email{priyank.du94@gmail.com}
\affiliation{Centre for High Energy Physics, Indian Institute of Science, C.\,V.\,Raman Avenue, Bengaluru 560012, India}
\affiliation{Department of Physics \& Astronomy, University of Southern California, Los Angeles, CA, 90007, USA}

\author{Ranjan Laha\orcidD{}}
\email{ranjanlaha@iisc.ac.in}
\affiliation{Centre for High Energy Physics, Indian Institute of Science, C.\,V.\,Raman Avenue, Bengaluru 560012, India}

	\date{\today}
	
	
	\begin{abstract}

A very pressing question in contemporary physics is the identity of Dark Matter (DM). Primordial Black Holes (PBHs) are one of the most well-motivated DM candidates. Light PBHs have been constrained by either the non-detection of their Hawking radiation itself, or by the non-observation of any measurable effects of this radiation on astrophysical and cosmological observables. We constrain the PBH contribution to the DM density by non-detection of their Hawking radiation's effect on the intergalactic medium (IGM) temperature evolution. We use the latest deductions of IGM temperature from Lyman-$\alpha$ forest observations. We put constraints on the fraction of DM as PBHs with masses $5 \times 10^{15}$\,g – $10^{17}$\,g, separately for spinning and non-spinning BHs. We derive  constraints by dealing with the heating effects of the astrophysical reionization sources on the IGM in two ways. In one way, we completely neglect this heating due to astrophysical sources, thus giving us weaker constraints, but completely robust to the reionization history of the universe. In the second way, we utilise some modelling of the ionization and temperature history, and use it to derive more stringent constraints. We find that for non-spinning PBHs of mass $10^{16}$\,g, the current measurements can constrain the PBH-density to be $\lesssim$ 0.1\% of the total DM. We find that these constraints are competitive, and hence provide a new observable to probe the nature of PBH DM. The systematics affecting Lyman-$\alpha$ forest measurements are different from other constraining observations, and thus this is a complementary probe.
 
	\end{abstract}
	
	\maketitle
	
	\section{Introduction}
	\label{sec:introduction}

Several astrophysical and cosmological evidences point towards the existence of Dark Matter (DM), something that makes up around $\sim$ 85\% of matter content today. But all these evidences rely on the gravitational interactions of DM. Any non-gravitational nature of DM is yet to be discovered. Due to a wide variety of DM models and a huge range in the masses of possible DM candidates, there has been a plethora of research on the possibilities and methods of observing any imprints of non-gravitational DM interaction with Standard Model (SM) particles or with itself.\,\cite{Bertone:2004pz,Bertone:2016nfn,Cirelli:2024ssz,Planck:2018vyg}.

Primordial Black Hole\,\cite{Zeldovich:1967lct, Hawking:1971ei,1974Natur.248...30H, Hawking:1975vcx,Carr:1975qj} (PBH) is one of the most popular and widely discussed DM candidates\,\cite{Carr:2020xqk, Green:2020jor,Escriva:2022duf}. Various production mechanisms for PBH DM have been proposed in the literature and various aspects concerning their formation have been studied\,\cite{Shibata:1999zs,Harada:2013epa,Musco:2023dak,Niemeyer:1999ak,Musco:2018rwt,Hawking:1971ei,Carr:1974nx,Yoo:2020lmg,Clesse:2015wea,Jedamzik:1996mr,Harada:2016mhb,Green:2000he,Polnarev:1988dh,Bhaumik:2019tvl,Bhaumik:2022zdd,Chakraborty:2022mwu,Liu:2021svg,Escriva:2021aeh,Harada:2024jxl,Bhattacharya:2019bvk,Bhattacharya:2021wnk,Banerjee:2023qya,Afzal:2024xci,ShamsEsHaghi:2022azq,Cai:2023uhc,Huang:2023chx,Khlopov:2008qy,Belotsky:2014kca,Ai:2024cka}. Irrespective of its production mechanism, a PBH would emit Hawking radiation, the spectrum of which depends on the BH mass, charge, and spin\,\cite{Hawking:1975vcx, Page:1976df,Page:1976ki,Page:1976jj,MacGibbon:1990zk,Arbey:2019mbc,Auffinger:2022khh}.
Several terrestrial and space-based  telescopes have been used to directly search for Hawking evaporation from PBH DM. These efforts have resulted in very stringent constraints on the abundance of low-mass ($\sim$ $10^{15}$\,g - $5\times 10^{17}$\,g\,) PBH DM\,\cite{ Carr:2009jm, PhysRevLett.122.041104, Laha:2019ssq, DeRocco:2019fjq,  PhysRevLett.126.171101,Laha:2020ivk, Dasgupta:2019cae, PhysRevD.101.023010,Chen:2021ngo, Laha:2020vhg, Kim:2020ngi, Chan:2020zry, Berteaud:2022tws}. Besides these direct searches for Hawking evaporation from PBH DM, a number of techniques have also been explored to search for indirect imprints of this radiation. SM particles emitted by evaporating PBH DM can dump energy into their surroundings. Effects of this exotic energy injection have been searched using various cosmological observables like the Cosmic Microwave Background\,\cite{Clark:2016nst, Acharya:2020jbv,Cang:2020aoo,Yang:2022nlt} and the global 21 cm signal\,\cite{Clark:2018ghm, Mittal:2021egv, Natwariya:2021xki, Saha:2021pqf,Cang:2021owu,Halder:2021rbq,Mukhopadhyay:2022jqc}. 

Lyman-$\alpha$ ($\sim$1216 \AA) forest, which is a series of absorption lines seen in the spectra of distant quasars, remains one of the most important tracers of the diffuse inter-galactic medium (IGM)\,\cite{Walther:2018pnn,Boera:2018vzq,Gaikwad:2020art,Spina:2024uyc,McQuinn:2015icp,2010gfe..book.....M}. Besides being sensitive to the values of various cosmological parameters and the nature of reionizing sources\,\cite{viel2009cosmological,viel2010effect,Fernandez:2023grg, Ivanov:2024jtl,Seljak:2006bg, Bagherian:2024obh,Palanque-Delabrouille:2015pga, Gerardi:2022ncj,DESI:2024lzq, Cuceu:2022wbd, Santos:2003pc, Furlanetto:2004jz, Loeb:1999er, tillman2023exploration,Zhu:2024huz,Cain:2024fbi}, observations of these Lyman-$\alpha$ absorption features are sensitive to the thermal and ionization history of the IGM\,\cite{Walther:2018pnn,Boera:2018vzq,Gaikwad:2020art,Spina:2024uyc,2024arXiv240906405J,10.1093/mnras/266.2.343,Theuns_2002,Hui:2003hn, Puchwein:2014zsa,Sanderbeck:2015bba}. With the current and promising progress in Lyman-$\alpha$ forest observations, it is important to investigate the potential of these observations to uncover Beyond the Standard Model (BSM) physics. Several works have already demonstrated the power of Lyman-$\alpha$ forest in searching for various BSM scenarios\,\cite{Villasenor:2022aiy, Murgia:2019duy, Garzilli:2018jqh, Fuss:2022zyt, Wang:2013rha, Hooper:2022byl, Irsic:2017yje, Liu:2020wqz, Bolton:2022hpt, McDermott:2019lch, Witte:2020rvb, Munoz:2017qpy,2017MNRAS.471.4606A,Kobayashi:2017jcf,Rogers:2020ltq}. In this work, these observations are used for the first time to put limits on the abundance of low-mass evaporating PBH DM.

This paper is structured as follows. Section (\ref{PBH review}) gives an overview of the energy injection into the IGM by evaporating PBHs. In section (\ref{history}), we briefly describe the evolution of the IGM temperture and ionization history in the presence of Hawking evaporating PBHs.  The first part of section (\ref{Lyman data}) is a brisk tour of the Lyman-$\alpha$ forest observations in the context of IGM temperature measurements. Then it lists the various datasets of temperature measurements using this technique that we make use of. We discuss our results in section (\ref{results}) and then conclude in section (\ref{conclusion}).

\section{Energy injection by evaporating PBH DM}
\label{PBH review}
A PBH with a given mass and angular momentum will evaporate via Hawking radiation \cite{Hawking:1975vcx}. The temperature of an uncharged spinning PBH is,
\begin{eqnarray}
T_{\rm PBH}=\frac{1}{4\pi G M_{\rm PBH}}\left(\frac{\sqrt{1-a_*^2}}{1+\sqrt{1-a_*^2}}\right),
\label{eqn:1}
\end{eqnarray}
where 
\begin{eqnarray}
a_*=\frac{J}{G M_{\rm PBH}^2},
\end{eqnarray}
with $a_*$, $J$, and $M_{\rm PBH}$ being the reduced
spin parameter, magnitude of angular momentum, and the mass of the PBH, respectively, and $G$ is the gravitational constant. For a non-spinning PBH ($a_*=0$), the above expression simplifies to,
\begin{eqnarray}
T_{\rm PBH}=\frac{1}{8\pi G M_{\rm PBH}}=1.06\left(\frac{10^{13} \,\text{g}}{M_{\rm PBH}}\right) \text{GeV}.
\end{eqnarray}
The number of $i^{\rm th}$ particle, $N_i$ emitted per unit energy per unit time by a Hawking evaporating BH is given by\,\,\cite{Page:1976df,Page:1976ki,MacGibbon:1990zk,Arbey:2019mbc}
\begin{eqnarray}
\frac{d^2N_i}{dE dt}=\frac{1}{2 \pi}\sum_{dof}\frac{\Gamma_i(E,M_{\rm PBH},a_*,\mu)}{e^{E'/T}- (-1)^{2s}},
\end{eqnarray} 
with
\begin{eqnarray}
    E'=E-m\Omega, \,\,\,\,\, \Omega=\frac{a_*}{2 M_{\rm PBH} (1+\sqrt{1-a_*^2})},
\end{eqnarray}
where $\Gamma_i$ are the greybody factors\,\cite{Page:1976df,Page:1976ki,MacGibbon:1990zk} encoding the probability of an emitted particle overcoming the gravitational potential of the black hole so as to escape away from the black hole, and $E$ is the total energy of the emitted particle, $m$ is the axial quantum number, $s$ is the spin of the emitted particle, $\mu$ is the rest mass of the emitted particle, and $dof$ stands for the degrees of freedom associated with the emitted particle. 

\begin{figure}
	\begin{center}
	\includegraphics[height=7.2cm]{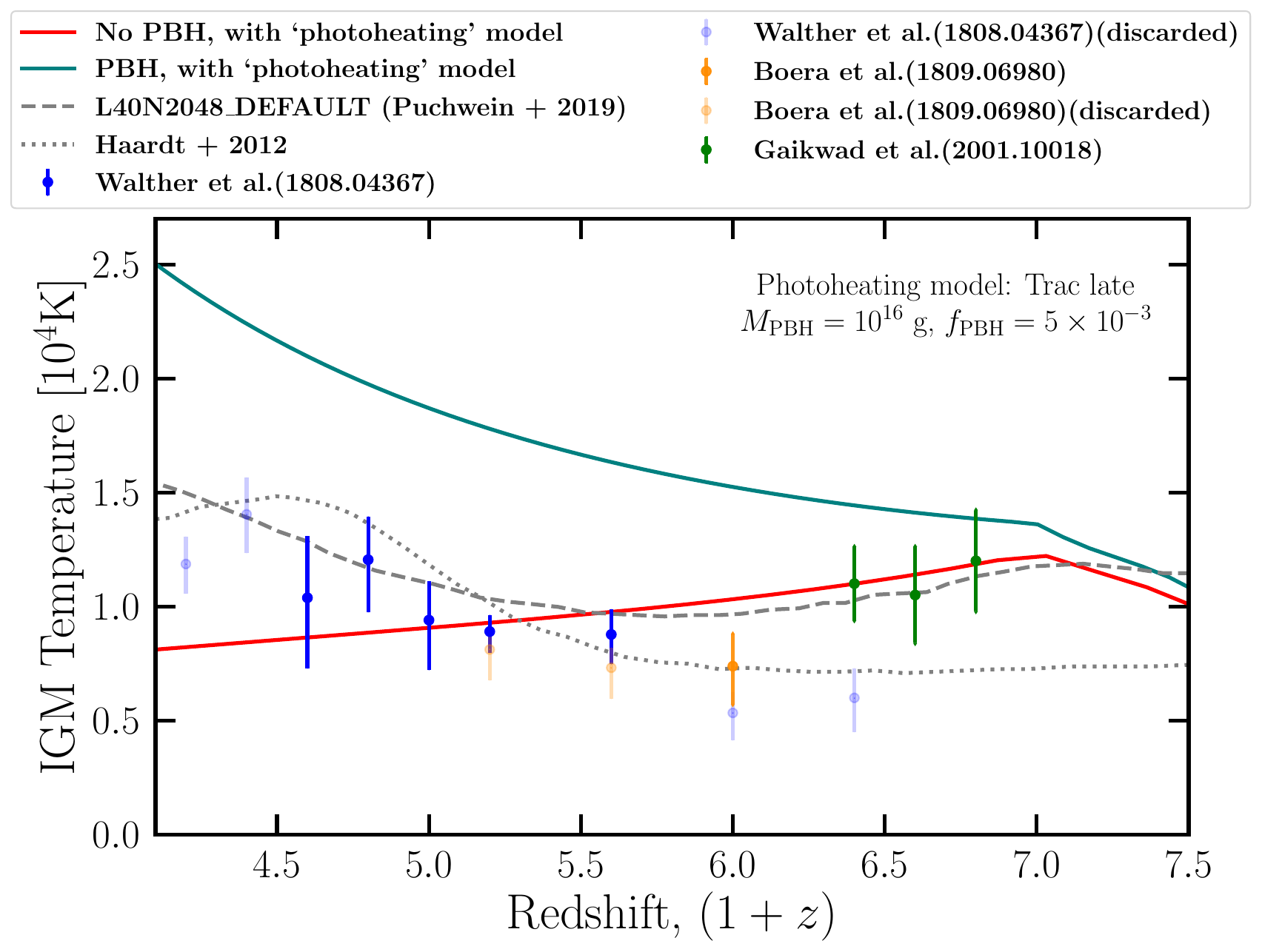}
	\caption{Temperature evolution of IGM as a function of redshift, in presence of non-spinning PBH DM (teal) and in the absence of PBH DM (red). In both these cases, the `Trac late' reionization model \cite{Trac:2018jla,Glazer:2018fnz} is considered (see text for more details). For comparison we show the Lyman-$\alpha$ temperature measurements from Walther et al.\,\cite{Walther:2018pnn} (blue), Boera et al.\,\cite{Boera:2018vzq} (orange), and Gaikwad et al.\,\cite{Gaikwad:2020art} (green). The deep shaded data-points are used in our analysis. The light shaded data-points not used either to avoid the second Helium reionization or to derive conservative limits on PBH DM. The grey dashed (Puchwein et al.\,\cite{Puchwein:2018arm}) and dotted (Haardt et al.\,\cite{haardt2012radiative}) lines correspond to two reionization models that are broadly in agreement with these datasets.}
	\label{fig:-IGMtemp} 
 \end{center}
\end{figure}
 \begin{figure*}[!htbp]
	\begin{center}
		\includegraphics[height=8.1cm]{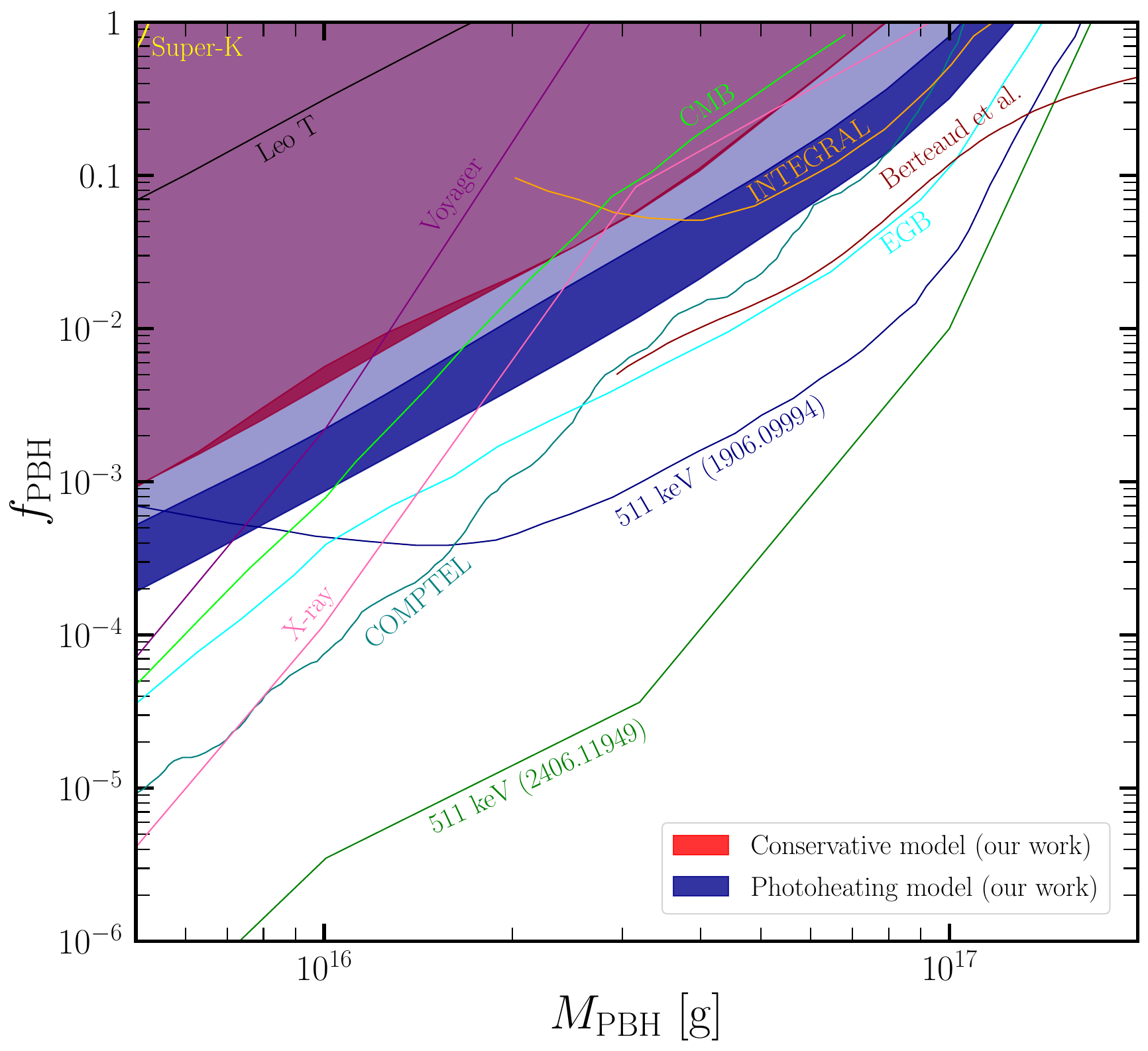}~~
		\includegraphics[height=8.1cm]{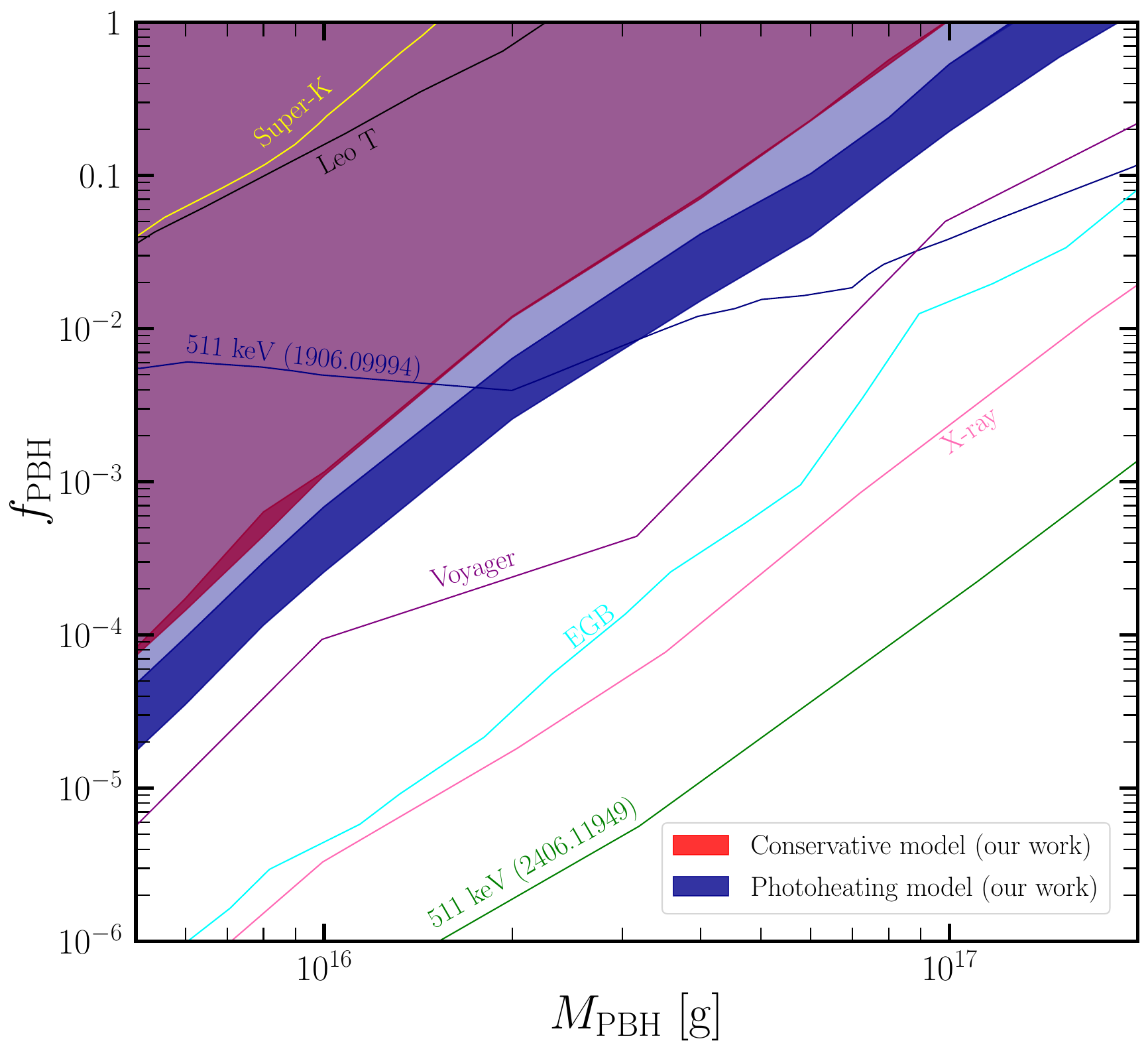}~~\\	
		\caption{Constraints on non-spinning ($a_*$ = 0) (left panel) and spinning ($a_*$ = 0.9999) (right panel) PBH DM (monochromatic mass distribution). Constraints from the conservative model are shown in red (deep red and the shaded region above that), whereas the photoheating model limits are shown in blue (deep blue and the shaded region above that). The deep blue and deep red bands are obtained by varying over the photoheating models used in this work (FlexKnot\,\cite{Planck:2018vyg}, Tanh\,\cite{Planck:2018vyg}, and Trac\,\cite{Trac:2018jla,Glazer:2018fnz}). Previous constraints include low-energy positron measurements from Galactic Centre by INTEGRAL (navy \cite{Laha:2019ssq} and green \cite{DelaTorreLuque:2024qms}), cosmic-ray flux measurement by Voyager (purple) \cite{DelaTorreLuque:2024qms}, INTEGRAL (orange, dark red) and COMPTEL (teal) measurements of the Galactic gamma-ray flux \cite{Laha:2020ivk,Berteaud:2022tws,PhysRevLett.126.171101}, Super-Kamiokande limit for diffuse supernovae neutrino background (yellow)\,\cite{Dasgupta:2019cae}, measurement of extra-Galactic gamma-ray emission  (aqua) \cite{PhysRevD.101.023010,Chen:2021ngo},  gas heating in Leo T (black) \cite{Laha:2020vhg,Kim:2020ngi}, PLANCK measurement of CMB (lime) \cite{Clark:2016nst}, and XMM-NEWTON measurement of Galactic  diffuse X-ray emission (pink) \cite{DelaTorreLuque:2024qms}. Note that two of the bounds - namely the cosmic-ray bounds from COMPTEL and INTEGRAL - for non-spinning PBH DM are not depicted for spinning PBH DM simply because the latter have not been derived in the literature.}
		\label{fig: -BoundLyman}
	\end{center}	
\end{figure*}
In this work, the Hawking evaporation spectra are obtained using the publicly available code, {\tt BlackHawk v2.2}\,\cite{Arbey:2019mbc,Arbey:2021mbl}. We have explicitly matched the spectra from {\tt BlackHawk} with ref.\,\cite{Page:1976df}. At low energies, for computing secondary photon and electron-positron spectra, {\tt BlackHawk} uses public {\tt Python} code {\tt Hazma}\,\cite{Coogan:2019qpu}. In this work we have used the total emission spectra (sum of primary and secondary) from PBHs in the mass range $\sim \, 5 \times 10^{15}\, {\rm g} -10^{17}$\,g. The lower limit on the mass of the PBHs comes from the fact that PBHs lighter than this mass lose a significant fraction of their mass in all the cosmological time from their formation epoch till today. Thus, their mass cannot be assumed to be constant, and thus, the emission spectra from them is time-dependent. We have probed those PBHs that have contributed the same to the dark matter density since their formation. These PBHs' mass doesn't change, and thus the evaporation spectrum of each emitted particle also remains the same throughout. Note that the only species of particles whose spectra we had to consider were photons and $e^{\pm}$. This is because only these particles are capable of efficiently interacting with the gas in the IGM and substantially affecting its thermal and ionization state\,\cite{Slatyer:2009yq,Marco_Cirelli_2011}. Other stable particles that are abundantly present in the Hawking radiation emitted by black holes of these masses are  neutrinos and anti-neutrinos, and these have no effect on the temperature evolution of the IGM due to their weak-interaction.

The rate of deposited energy into the IGM is related to the rate of energy injected from Hawking evaporation by\,\cite{Liu:2018uzy}
\begin{eqnarray}
\left(\frac{dE}{dVdt}\right)_{\rm dep}=f_c(z)\left(\frac{dE}{dVdt}\right)_{\rm inj},
\end{eqnarray}
where $dV$ is the volume element and $f_c(z)$ is the deposition efficiency into channel $c$ at a given redshift $z$\,\cite{PhysRevD.94.063507,PhysRevD.95.023010,Slatyer:2015kla}, respectively. The relevant channels are hydrogen ionization ($f_{ion}$), helium ionization ($f_{He}$), hydrogen excitation ($f_{exc}$), and heating of IGM ($f_{heat}$). The $f_c(z)$'s have been computed in refs.\,\cite{PhysRevD.94.063507,PhysRevD.95.023010,Slatyer:2015kla} and have been implemented in the publicly available code {\tt DarkHistory}\,\cite{Liu:2019bbm, Liu:2020wqz, Liu:2023fgu,Liu:2023nct,Qin:2023kkk,Sun:2023acy,Xu:2024vdn}. In our work we have used {\tt DarkHistory} to calculate the evolution of IGM temperature and ionization in presence of evaporating PBHs. {\tt Darkhistory} enables us to calculate the the modified thermal and ionization state of the IGM self-consistently. What this means is that the effect of a modified ionization history at any redshift $z$ due to evaporating PBHs affects the heat deposition rates into the IGM also, and vice-versa. We have computed both the modified ionization history and the temperature history fully taking into account each other's evolution (called backreaction in \cite{Liu:2020wqz}.)

We can write the energy deposited per unit volume per unit time from a PBH of mass $M_{\rm PBH}$,  as,
\begin{eqnarray}
\left(\frac{dE}{dVdt}\right)_{\rm dep}=\left(\frac{dE}{dVdt}\right)_{\rm dep}^\gamma + \left(\frac{dE}{dVdt}\right)_{\rm dep}^{e^\pm},
\end{eqnarray}
with
\begin{eqnarray}
\left(\frac{dE}{dVdt}\right)_{\rm dep}^\gamma =  \int\int f_c(E_\gamma,z)E_\gamma\left(\frac{d^2N}{dE dt}\right)_\gamma \nonumber\\
\times n_{\rm PBH}(M_{\rm PBH})\,\Psi(M_{\rm PBH})\,dM_{\rm PBH}\,dE_\gamma,
\end{eqnarray} 
and
\\
\begin{align}
&\left(\frac{dE}{dVdt}\right)_{\rm dep}^{e^\pm}
=\int\int 2f_c(E_e-m_ec^2,z)(E_e-m_ec^2)\nonumber\\
&\times \left(\frac{d^2N}{dE dt}\right)_{e^\pm}n_{\rm PBH}(M_{\rm PBH})\,\Psi(M_{\rm PBH})\,dM_{\rm PBH}\,dE_e.
\end{align} 
Here, $n_{\rm PBH}$ and $\Psi(M_{\rm PBH})$ are the number density and mass distribution function of PBHs, respectively. The number density of PBH DM, $n_{\rm PBH}$ can be written as
\begin{eqnarray}
n_{\rm PBH}=f_{\rm PBH}\frac{\rho_c\,\Omega_{\rm DM}}{M_{\rm PBH}}(1+z)^3,
\label{eq: number density PBH}
\end{eqnarray}
with $f_{\rm PBH}$, $\rho_c$, and $\Omega_{\rm DM}$ being the fraction of DM in the form of PBHs, the critical density, and DM density parameter in the present universe.
In our work we have assumed monochromatic mass distributions of PBH DM,  i.e., all PBHs are assumed to be of the same mass, but the value of this mass is varied. Thus, $\Psi(M_{\rm PBH})$ takes the form of a dirac delta function peaked at a variable PBH mass $M_{\rm PBH}$.

\section{Ionization and Thermal history of the IGM}
\label{history}
Having discussed the deposition of the injected energy sourced by Hawking evaporation of PBHs into the IGM, we
now describe the effects of this deposited energy on the IGM temperature and the ionized hydrogen fraction. The rate of change of the IGM temperature $(\dot{T_m}$) is given by\,\cite{Peebles:1968ja, 1969JETP...28..146Z, Ali-Haimoud:2010hou, Liu:2019bbm},
\begin{eqnarray}
\dot{T}_\text{m} = \dot{T}_\text{m}^{(0)}+\dot{T}_\text{PBH} \,,
\label{eq:temp_diff_eq}
\end{eqnarray}
where the first term on the right hand side describes contributions due to all standard astrophysical and cosmological processes. The second term represents the contribution due to PBH energy injection. The term $\dot{T}_\text{m}^{(0)}$ is,
\begin{eqnarray}
\dot{T}_\text{m}^{(0)} = \dot{T}_\text{adia} + \dot{T}_\text{C} + \dot{T}_\text{atom} + \dot{T}^\star \,.
\label{eq:temp_diff_eq2}
\end{eqnarray}
In the above equation, $\dot{T}_\text{adia}$ and $\dot{T}_\text{C}$ are the IGM cooling rate due to adiabatic expansion of the universe and heating/\,cooling due to (inverse) Compton scattering, respectively. These terms can be written as,
\begin{eqnarray}
\dot{T}_\text{adia} = -2 H T_\text{m} \,,\nonumber \\
\dot{T}_\text{C} = -\Gamma_C (T_\text{CMB} - T_\text{m}) \, ,
\end{eqnarray}
where,
\begin{eqnarray}
\Gamma_C = \frac{x_\text{e}}{1 +f_{\rm He} + x_\text{e}} \frac{8 \sigma_T a_r T_\text{CMB}^4}{3 m_\text{e}} \,.
\end{eqnarray}
In the above equations, $H$ is the Hubble parameter, $T_\text{CMB}$ is the Cosmic Microwave Background (CMB) temperature, $x_e$ is the free electron fraction ($\equiv n_\text{e}/{n_\text{H}}$). The parameter, $ f_{\rm He} \equiv n_\text{He}/{n_\text{H}}$, where  $n_\text{He}$ and $n_\text{H}$ are the number densities of helium and hydrogen nuclei, respectively; $\sigma_T$ is the Thomson cross section, $a_r$ is the radiation constant, and $m_e$ is the electron mass.
The term
$ \dot{T}_\text{atom} $ in Eq.~\eqref{eq:temp_diff_eq2} represents the cooling term due to all atomic processes like collisional ionization, collisional excitation and Coulomb heating\,\cite{PhysRevD.94.063507}. Finally,  the $\dot{T}^\star$ term in Eq.~\eqref{eq:temp_diff_eq2} is the photoheating term that accounts for the heating due to photoionization from various astrophysical sources. This term is dictated by the ionization history of the IGM, which we briefly describe in what follows.

The evolution equation for the ionized hydrogen fractionin the IGM is:
\begin{eqnarray}
\dot{x}_\text{HII} = \dot{x}_\text{HII}^{(0)}+ \dot{x}_\text{HII}^\text{PBH}\,,
\label{eq:ionization_diff_eq}
\end{eqnarray}
where $x_\text{HII}=n_\text{HII}/{n_\text{H}}$, with $n_\text{HII}$ being the number density of ionized hydrogen. The first term on the right hand side is again the contribution due to the standard astrophysical sources and the second term represents the contribution due to PBH energy injection. The first term in
Eq.\,\eqref{eq:ionization_diff_eq} is: 
\begin{eqnarray}
\dot{x}_\text{HII}^{(0)} = \dot{x}_\text{HII}^\text{atom}  + \dot{x}_\text{HII}^\star\,.
\end{eqnarray}
where the first term on the right hand side is due to the atomic processes such as collisional ionization and recombination\,\cite{Peebles:1968ja, 1969JETP...28..146Z, Seager:1999bc,Seager:1999km,Chluba:2010ca,Ali-Haimoud:2010hou,Liu:2019bbm}. The second term, $\dot{x}_\text{HII}^\star$ takes into account the photoionization due to astrophysical sources, which can be calculated given a reionization model. However, following Refs.~\cite{Liu:2020wqz}, we  instead take the model independent approach of using the Planck CMB constraints on free electron fraction ($x_e$) to fix $\dot{x}_\text{HII}^\star$,\begin{eqnarray}
\dot{x}_\text{HII}^\star = \left(\frac{\dot{x}_\text{e}^\text{P}}{1 + f_{\rm He}} - \dot{x}_\text{HII}^\text{atom} - \dot{x}_\text{HII}^\text{PBH}\right) \,.
\label{eq:photoionization_rate}
\end{eqnarray}
Here, $x_\text{e}^\text{P}$ is taken from the Planck 95\% confidence region limits on free electron fraction using `Tanh' and `FlexKnot' models\,\cite{Planck:2018vyg}. In addition, we also use the 95\% confidence region limit from parametrization described in Refs.\,\cite{Trac:2018jla,Glazer:2018fnz}, hereafter referred to as the `Trac' model.
We note that, at sufficiently high redshifts\,\footnote{Please refer to Ref.\,\cite{Liu:2020wqz} for more details on how to fix the value of redshift above which $\dot{x}_\text{HII}^\star = 0$. $\dot{x}_\text{HII}^\star$ is taken to be zero.}, $\dot{x}_\text{HII}^\star =0$.

Finally, we use the following parametrization to fix $\dot{T}^\star$,
\begin{alignat}{1}
    \dot{T}^\star = \begin{dcases}
        \dot{x}_\text{HII}^\star (1 + f_{\rm He}) \Delta T \,,\,\,\,\, & x_\text{HII} < 0.99 \,, \\
        \sum_{\text{X}=\{ \text{H}, \text{He} \}} \frac{h\nu_\text{X}}{3(\gamma_\text{X} - 1 + \alpha_\text{bkg})} \alpha_{\text{A},\text{X}}\, n_\text{X} \,, & x_\text{HII} \geq 0.99 \,,
    \end{dcases}
    \label{eq:modeled_term}
\end{alignat}
In the above equation, $\Delta T$ is a free parameter whose value is known to lie in the range $2\times10^4\,{\rm K} - 3\times10^4$\,K as discussed in Refs.\,\cite{Onorbe:2016hjn,McQuinn:2015gda,10.1093/mnras/266.2.343,mcquinn2012constraints,Sanderbeck:2015bba}. We shall use the value of $\Delta T$ within this range which fits our temperature measurement data the best. Furthermore, $\nu_\text{X}$ is the frequency corresponding to the ionization potential of species $\text{X}\in\{ \text{HI}, \text{HeI} \}$, $h$ is Planck's constant, $\gamma_\text{X}$ is the index for photoionization cross-section, $\alpha_\text{bkg}$ is the index for power-law ionizing background\,\cite{McQuinn:2015gda,Sanderbeck:2015bba,Puchwein:2014zsa}, and $\alpha_{\text{A},\text{X}}$ are the case-A recombination factors associated with the transitions XI $\rightarrow$ X\,\cite{Sanderbeck:2015bba}. Like $\Delta T$, $\alpha_{\rm bkg}$ is a free parameter which can be varied to get a best-fit to the temperature data, and whose value is also known to lie within a range- $-0.5<1.5$ in this case.

We have considered two models, namely `conservative' and `photoheating', to take into account effects of reionization on IGM temperature. For our `conservative' model, we assume $\dot{T}^\star=0$, i.e., there is no photoheating due to reionizing sources. In realistic scenarios there is always photoheating present during reionization, and thus the `conservative' approach gives us robust limits. In case of `photoheating' model we assume that $\Delta T\geq2\times10^4$\,K, and derive the corresponding bounds on PBH DM.

The effects of evaporating PBHs on the temperature and ionization evolution are given as:
\begin{eqnarray}
\dot{T}_\text{PBH} =\frac{2 f_{heat}(z)}{3(1+f_{\rm He}+x_{e})n_{\text{H}}}\left(\frac{dE}{dVdt}\right)_{\rm inj}\nonumber\,,\\
\dot{x}_\text{HII}^\text{PBH}=\left[\frac{f_{ion}(z)}{\mathcal{R}n_{\text{H}}}   +\frac{(1-\mathcal{C})f_{exc}(z)}{0.75\mathcal{R}n_{\text{H}}}       \right]\left(\frac{dE}{dVdt}\right)_{\rm inj}\,,
\label{eq:temp_ion_PBH}
\end{eqnarray}
where $\mathcal{R} = 13.6$ eV is the hydrogen ionization potential and $\mathcal{C}$ is the Peebles' C factor\,\cite{Peebles:1968ja, Ali-Haimoud:2010hou}.

\section{Lyman-$\alpha$ Forest as a probe of IGM temperature}
\label{Lyman data}
Lyman-$\alpha$ transitions occur in the hydrogen atom when the electron is excited from the ground state ($n=1$) to the first excited state ($n=2$) and vice versa. Thus, whenever photons with wavelength $\sim$1216 \AA, which corresponds to the energy difference between the ground state and first excited state, encounter neutral hydrogen they are absorbed along the line of sight. Photons coming towards us from a distant quasar, encounter neutral hydrogen in IGM at different redshifts, and get absorbed. 
Due to the expansion of the Universe, these transitions lead to a series of absorption lines in the observed quasar spectrum. The resulting spectrum resembles a forest, and is thus called Lyman-$\alpha$ forest. 
The Lyman-$\alpha$ absorption is quantified by the optical depth ($\tau$) along the line of sight, given by\,\cite{1965ApJ...142.1633G},
\begin{eqnarray}
\tau_{\mathrm{Ly}}(z)=1.3 \Delta_b\left(\frac{x_{\mathrm{HI}}}{10^{-5}}\right)\left(\frac{1+z}{4}\right)^{3 / 2} \nonumber\\
\times \left(\frac{d v / d x}{H(z) /(1+z)}\right)^{-1},
\end{eqnarray}
where $z$ is the redshift of the gas at fixed density with a line of sight gradient, $dv / dx$ , $\Delta_b$ is the baryonic density in units of cosmic mean density, $\Delta_b= \frac{\rho_b}{\langle\rho_b\rangle}$, $\rho_b$ being the baryonic density. In the above equation, $x_{\mathrm{HI}}$ is the neutral hydrogen fraction. Using the `fluctuating 
Gunn-Peterson approximation'\,\cite{ Weinberg:1997zx,Croft:1997jf}, peculiar velocity gradient, $dv / dx\sim H(z) /(1+z)$.

During hydrogen reionization, IGM gets ionized and then photoheated due to ionization fronts from different reionizing sources. The exact temperature of IGM during this time depends on the incident spectra and velocity of the ionization front \,\cite{10.1093/mnras/266.2.343,McQuinn:2015icp}. After the completion of reionization, the IGM is cooled mainly due to cosmic expansion and Compton cooling off CMB\,\cite{McQuinn:2015gda}. The IGM temperature is assumed to be a power law function of gas density\,\cite{Hui:1997dp},
\begin{eqnarray}
T(\Delta_b)=T_0 \times \Delta_b^{\gamma-1},
\end{eqnarray}
where, $T_0$ is the IGM temperature at mean density $\langle\rho_b\rangle$ and $\gamma$ is the power-law index of this relation. During reionization, $\gamma$ becomes close to 0 or negative\,\cite{Trac:2008yz,furlanetto2009equation}, whereas long after reionization,  $\gamma\sim 1.6$\,\cite{Rudie2012THETR,Bolton:2013cba}.

The IGM temperature at a particular redshift contributes to the thermal motion of the neutral hydrogen atoms in the IGM. Due to this thermal motion, the Lyman-$\alpha$ absorption features also show a cumulative Doppler shift, known as Doppler broadening. As a result, widths of the observed Lyman-$\alpha$ forest absorption features are affected. Lyman-$\alpha$ forest thus acts as a thermometer for the IGM. In the presence of any excess heat injection, the observed absorption features will be wider than expected due to standard astrophysical contributions\,\cite{Garzilli:2015bha} (broadening due to gas pressure is sub-dominant\,\cite{peeples2010pressure}).

From measurements of both $T_0$ and $\gamma$ as a function of redshift, one can know about the temperature evolution of the IGM during and after hydrogen and helium reionization\,\cite{10.1093/mnras/266.2.343,Theuns_2002,Hui:2003hn, Puchwein:2014zsa,Sanderbeck:2015bba,Gaikwad:2020art}. High redshift quasar spectra are compared with mock Lyman-$\alpha$ forest blue{spectra} generated by hydrodynamical simulations, to infer these parameters. These inferences are based on the Voigt profile decomposition, flux power spectrum, probability distribution of flux.\,\cite{Schaye:1999at,Rudie2012THETR, Bolton:2013cba,rorai2018new,Hiss:2017qyw,             Theuns:1999mz,bolton2008possible,viel2009cosmological,calura2012lyman,lee2015igm,rorai2017exploring,Garzilli:2015iwa,walther2017new,Boera:2018vzq,khaire2019power,Walther:2018pnn,Gaikwad:2020art,Irsic:2023equ}. Most of these measurements use complementary methods. Below we briefly talk about the datasets that we use in this work:

\subsection*{Walther et al. (2019)\,\,\cite{Walther:2018pnn}}
Lyman-$\alpha$ flux power spectrum probes a large scale starting from $\sim$ 500 kpc to $\sim$ 10 Mpc. The shape of the power spectrum and the small scale cutoffs help in breaking degeneracies between thermal parameters like $T_0$ and $\gamma$. Ref.\,\cite{Walther:2018pnn} used high resolution quasar spectra from surveys like BOSS\,\cite{BOSS:2013rpr}, XQ-100\,\cite{Irsic:2017sop}, Keck/HIRES\,\cite{OMeara:2015lwd,o2017second}, Magellan/MIKE\,\cite{Viel:2013fqw} and compared the resulting power spectra with those obtained from hydrodynamical simulation\,\cite{Almgren:2013sz}, to obtain temperature measurements in the redshift range $1.8\,\textless\,z\,\textless\,5.2$. \subsection*{Boera et al. (2019)\,\,\cite{Boera:2018vzq}} Ref.\,\cite{Boera:2018vzq} looked at high resolution spectra of 15 quasars obtained from Keck/HIRES\,\cite{Vogt:1994fao} and VLT/UVES\,\cite{2000SPIE.4008..534D}. Comparing the flux power spectra from observational data with predictions from a suite
of hydrodynamical simulations, they obtained the IGM temperature at mean density in the redshift range $4\,\textless\,z\,\textless\,5.2$.
\subsection*{Gaikwad et al. (2020)\,\,\cite{Gaikwad:2020art}} At $z \,\textgreater\,5$, the Lyman-$\alpha$ optical depth increases with redshift. As a result, the transmitted Lyman-$\alpha$ flux becomes close to zero with few transmission spikes. These spikes correspond to underdense, highly ionized gas\,\cite{Gnedin:2016wrw,Chardin:2017fbm,Nasir:2019iop}. Ref.\,\cite{Gaikwad:2020art} investigated the dependence of these spike widths on the gas temperature. By comparing spectra from five QSOs (Quasi-Stellar Objects) at $z \,\textgreater\,6$ with hydrodynamical simulations, the authors obtained the gas temperature at mean density in the redshift range $5.4\,\textless\,z\,\textless\,5.8$. 
\vspace{20px}

We are not using data points below redshift of 3.6  in order to avoid second helium reionization.  {\tt DarkHistory} uses transfer functions with $x_{\rm HeIII}=0$. Thus we derive conservative limits by avoiding HeII reionization and the corresponding heating of IGM. We note that, for redshifts 4.2, 4.6, and 5, Walther et al.\,temperature measurements agree with Boera et al., within the error bars. For redshifts 4.2 and 4.6, we use Walther et al.\,\,data points, whereas for redshift 5, we use data from Boera et al. This is done to choose higher temperature data points which in turn results in conservative limits on PBH DM. For the same reason we choose the temperature measurement in the first bin of Gaikwad et al., instead of the last data point of Walther et al.

\section{results}
\label{results}
For each $M_{\rm PBH}$ and $f_{\rm PBH}$, we obtain the temperature and ionization evolution as a function of redshift for both `conservative' and `photoheating' models. One such example is shown in Fig.\,\ref{fig:-IGMtemp}. Here we show the IGM temperature evolution in presence of evaporating PBH and `Trac late' photoheating model. For comparison, we also show the datasets (deep shaded points) from\,\cite{Walther:2018pnn,Boera:2018vzq,Gaikwad:2020art}, used in this work. Given the datasets, for the `Trac late' model we find a best fit temperature evolution (shown in red line) with $\Delta T=2\times10^4$\,K and $\alpha_{\rm bkg}=0.66$. One important feature to note here is that though both astrophysical sources and evaporating PBH DM can heat up the gas, their temperature evolution with redshift is much different. Hence, in future with more temperature measurements one can in principle discriminate PBH signal from that of standard astrophysics. In Fig.\,\ref{fig:-IGMtemp}, we also show some other background astrophysical models (grey dashed and dotted lines) from literature (taken from Fig.\,10 of \cite{Gaikwad:2020art}) that are broadly in agreement with the datasets used \cite{Puchwein:2018arm,haardt2012radiative}. This shows that given the error bars in the datasets and modelling of  astrophysical sources during reionization, there are many background models that are in agreement with the Lyman-$\alpha$ forest IGM temperature observations. We emphasize that we are not trying to fit the observed temperature measurements to those obtained from PBH DM. Instead using the temperature datasets, we try to constrain the existence of exotic heat injection sources like PBH DM on top of a known astrophysical model. In order to derive bounds on evaporating PBH DM, for `conservative' and `photoheating' models we follow the $\chi^2$ analysis described in Ref.\,\cite{Liu:2020wqz}, i.e., for `conservative' model we penalize the PBH DM contribution only when it exceeds the temperature measurements, whereas, for `photoheating' model we follow the standard $\chi^2$ analysis. 

In Fig.\,\ref{fig: -BoundLyman}, we show our constraints on evaporating PBH DM. In the left panel we show our bounds on non-spinning PBH DM for `conservative' (red line) and `photoheating' (blue line) models. The deep shaded regions show the variation in the constraints when we vary the electron fraction limits for `Tanh', `FlexKnot', and `Trac' models. The regions above the red and blue lines are ruled out by the datsets used in this work. For comparison we also show the combined previous limits on PBH DM from various different astrophysical and cosmological observables\,\cite{Laha:2019ssq,PhysRevLett.122.041104,PhysRevLett.126.171101,Laha:2020ivk,Dasgupta:2019cae,PhysRevD.101.023010,Chen:2021ngo,Laha:2020vhg,Kim:2020ngi,Clark:2016nst,Chan:2020zry}. We also show our bounds when the spin parameter of PBH is, $a_*=0.9999$, in the right panel of Fig.\,\ref{fig: -BoundLyman}. The corresponding previous limits on spinning PBH DM are also shown. As expected, we find that when we do not include any heating effects from reionization, we obtain less stringent constraints. But these constraints are completely robust to any astrophysical uncertainties related to the process of reionization. In the case of `photoheating' model, both photoionization and photoheating are included, which results in stronger constraints on PBH DM. Since the `photoheating' constraints are affected by photoheating during reionization, hence how reionization proceeds affects these constraints much more, and thus these constraints are represented by a wider band as compared to the `conservative' constraints.
In both non-spinning and spinning cases, we find that our limits are complementary to the limits present in the parameter space. Our limits are derived assuming a monochromatic mass distribution for PBH DM. We leave the exploration of the Lyman-$\alpha$ forest limits with other mass distributions (like log-normal distribution) for future work. 

For some other constraints exploiting the evaporation of PBHs, the reader is urged to consult \cite{Yang:2024vij,Iguaz:2021irx,Laha:2020ivk,Liu:2024kub,Tan:2024nbx,Su:2024hrp,Huang:2024xap,Tan:2023fdy,Melikhov:2023vto,Bernal:2022swt,Chen:2021ngo,Natwariya:2021xki,Mittal:2021egv,Korwar:2023kpy,Tan:2022lbm,Clark:2016nst,Cang:2021owu,Kim:2020ngi,Chan:2020zry,Liu:2023cqs}. QED corrections to Hawking radiation have been discussed in \cite{Silva:2022buk,Koivu:2024gjl,Vasquez:2024hpc}. Constraints on evaporating PBH DM that take into account a possible quantum gravity effect, termed `memory burden effect', can differ from those based on the standard Hawking radiation emission and are discussed in \cite{Thoss:2024hsr,Kohri:2024qpd,Dvali:2024hsb,Haque:2024eyh,Alexandre:2024nuo}. There have also been studies on PBH formation in alternate metrics\,\cite{Calza:2024fzo,Calza:2024xdh}. Heavier PBHs have also been constrained using various other techniques, a non-exhaustive assortment is \cite{Zhou:2021tvp,EROS-2:2006ryy,Niikura:2019kqi,Mroz:2024mse,Capela:2013yf,Pani:2014rca,Inoue:2017csr,Serpico:2020ehh,Nakama:2017xvq,Smyth:2019whb,Niikura:2017zjd,Katz:2019qug,Chen:2019xse,Franciolini:2021tla,Takhistov:2021aqx,Lu:2020bmd,Agius:2024ecw,Andres-Carcasona:2024wqk,Franciolini:2023opt,Ng:2022agi,DeLuca:2024uju}. We note in passing that PBHs in the mass window $10^{18}$ g to $10^{22}$ g have proven to be notoriously challenging to be probed. Some papers that discuss various ideas for probing PBHs of these masses are \cite{1992ApJ...386L...5G,Katz:2018zrn,Nemiroff:1995ak,Jung:2019fcs,Bai:2018bej,Laha:2018zav,Montero-Camacho:2019jte,Ray:2021mxu,Ghosh:2021gfa,Ghosh:2022okj,Tran:2023jci,Gawade:2023gmt,Tamta:2024pow,Crescimbeni:2024cwh,Thoss:2024dkg,Crescimbeni:2024qrq}.
\section{conclusions}
\label{conclusion}
In this work, we show the potential of probing evaporating PBH DM using existing Lyman-$\alpha$ forest observations. IGM temperature measurements from comparison of Lyman-$\alpha$ forest data and hydrodynamical simulations, are sensitive to any sort of exotic heat injection. Using the well known evaporation spectra from PBH DM and using the existing IGM temperature datasets, we put competitive limits on the abundance of PBH DM. For taking into account the effects of reionization, instead of using any particular reionization model, we rely on the Planck limits on free electron fraction. In our ‘conservative’ case, we neglect effects of photoheating to arrive at the most robust limits. In the `photoheating’ case, we take into account the effects of photoheating due to reionizing sources. As expected, our limits improve when we take into account the effects of photoheating.  In future, we will have more high resolution measurements of distant quasars. This will result in more accurate estimates of IGM temperatures, with smaller error bars. Besides, the limits on various reionization models are also expected to improve. As a result, Lyman-$\alpha$ forest observations in future will either discover evaporating PBH DM or put more stringent limits on the parameter space.

\section{Acknowledgements}
We thank Prakash Gaikwad for detailed discussions on different hydrodynamical simulations and temperature measurements in the context of Lyman-$\alpha$ forest. We thank Hongwan Liu for detailed help regarding {\tt DarkHistory}. We also thank Vikram Khaire, Girish Kulkarni, Aseem Paranjape, Joachim Kopp, Marco Cirelli, Felix Kahlhoefer, Alejandro Ibarra, Nirmal Raj, Aniket Joglekar, Wenzer Qin, and Pravabati Chingangbam for useful discussions. AKS and AS acknowledge the Ministry of Human Resource Development, Government of India, for financial support via the Prime Ministers’ Research Fellowship (PMRF), PP acknowledges the support from the National Post-Doctoral Fellowship by the Science and Engineering Research Board (SERB), Department of Science and Technology (DST), Government of India (SERB-PDF/2023/002356) and the IOE-IISc fellowship program, RL acknowledges financial support from the Infosys foundation (Bangalore), institute start-up funds, the Department of Science and Technology (Govt.\,of India) for
the grant SRG/2022/001125, and ISRO-IISc STC for the grant no.
ISTC/PHY/RL/499.

The first three authors have contributed equally to this work.

\appendix*
\section{Modifying \texttt{DarkHistory}}

As mentioned in Section (\ref{PBH review}), we used the publicly available code \texttt{DarkHistory} to compute the temperature and ionization-state evolution of the IGM, when energy is injected into it by the evaporating PBHs. \texttt{DarkHistory}, in its available form, can compute how dark matter particles of a given mass $m$ and  lifetime $\tau$ decaying into specified SM particles affect the temperature and ionization history of the IGM. {\texttt{DarkHistory} can also do the same for annihilating DM, but we don't make use of that functionality. We had to make some modifications to the code to be able to use it for the case of evaporating BHs.

 We modified the code in two different ways in order to cross-check our results. We get the same results either way and this ensures that our computations are trustworthy. We have also calculated the sensitivities on PBH DM from global 21 cm signal and the results match with ref.\,\cite{Clark:2018ghm}. We outline the two different ways of modifying the code below.

\subsection{Method A}
\label{sec: Method A}

In this method, we make use of the fact that evaporating PBHs can be thought of as decaying dark matter. A single PBH can be imagined to be a lumped aggregate of (imaginary) many dark matter particles which are decaying, leading to the evaporation of the PBH as a whole as they disappear upon decaying. To make use of this idea, we run the code for the case of a DM particle decaying into two SM particles. When the unmodified code is run for any such case, the spectra of $e^{\pm}$ and photons that are finally injected into the IGM for a given mass of $\chi$ are calculated from the PPPC4DMID\,\cite{Marco_Cirelli_2011} results. We change the code such that instead of calculating these spectra from the PPPC4DMID results, it takes as input some other spectra that we specify. The specified spectrum of photons, for instance, that replaces the photon spectrum from PPPC4DMID for one DM particle decaying, is chosen such that the composite spectrum of all the photons injected into the IGM (per unit volume per unit time) is the same as the spectrum of photons, that evaporating PBHs of a given mass would have injected. We do the same for electrons and positrons. Note that since the spectrum of electrons, positrons and photons injected by the decay of each $\chi$ are specified by us, we can run the code for any combination of arbitrary stable SM final states.

To achieve this, we have to calculate the spectrum of $e^{\pm}$ and photons that should be injected upon the decay of each $\chi$ particle. For concreteness, let us just focus on the case of photons as the case of $e^{\pm}$ is completely analogous.

We need to determine the spectrum of photons injected when the particle $\chi$, of mass $m$ and lifetime $\tau$ decays. First, what values of $m$ and $\tau$ should be chosen for our $\chi$? As it will become clear later, we can choose any `reasonable' values of these two parameters. By `reasonable', we mean values that are typical of DM particle candidates, so that the code works fine for these values. We choose $m=1$ GeV and $\tau=10^{25}$ s.

We now turn our attention to the meat of the matter. We want to know what spectrum of photons should be specified for injection upon decay of a single particle $\chi$, such that the total spectrum of all photons injected per unit volume is the same as if PBHs of mass $M_{\mathrm{PBH}}$ and spin parameter $a^*$ were evaporating. We know the number of photons of a given energy emitted by such PBH of per unit time, denoted by
\begin{eqnarray}
\left(\frac{d^2N}{dEdt}\right)_\gamma\,.
\label{eq:appendix_A_1}
\end{eqnarray}
Over a time period $\tau$, the PBH injects the following spectrum of photons:
\begin{eqnarray}
\tau\left(\frac{d^2N}{dEdt}\right)_\gamma \,.
\label{eq:appendix_A_2}
\end{eqnarray}
We want to achieve the injection of exactly this spectrum in time $\tau$, but from the decay of particles $\chi$. The spectrum injected by each particle should thus be:
\begin{eqnarray}
\frac{\tau}{n}\left(\frac{d^2N}{dEdt}\right)_\gamma,
\label{eq:appendix_A_3}
\end{eqnarray}
where $n$ is the number of particles $\chi$ that can take the place of PBHs. Since the density of dark matter is a fixed quantity, independent of whether it is composed of PBHs or particles $\chi$, $n$ is just $\frac{M_{\mathrm{PBH}}}{m}$. Thus, the spectrum that we have to specify for injection upon decay of each $\chi$ is given by
\begin{eqnarray}
\frac{\tau m}{M_{\mathrm{PBH}}}\left(\frac{d^2N}{dEdt}\right)_\gamma \,.
\label{eq:appendix_A_4}
\end{eqnarray}

This spectrum is what goes as the $\left(\frac{dE}{dVdt}\right)_{\rm{inj}}$ term in Eq.\,(\ref{eq:temp_ion_PBH}), and then \texttt{DarkHistory} takes over and computes the temperature and ionization history of the IGM.

Once we have the spectrum to be injected upon decay of each particle, there is one slight complication we have to deal with. \texttt{DarkHistory} can work with spectra only with a specific set of energy bins. Thus, one has to convert rebin spectra by applying the \texttt{.rebin(photeng)} method for the photon spectra and the \texttt{.rebin(eleceng)} for that of $e^{\pm}$.

It may come off as baffling that how this spectrum could depend on $\tau$ and $m$, whose values, as remarked above, are arbitrary. The way to unpack this is to notice that the rate of total number of dark matter particles decaying per unit volume by decaying particle dark matter is given by 

\begin{eqnarray}
\left(\frac{dN_\chi}{dVdt}\right)_{\mathrm{inj}}=\frac{\rho_\chi}{m\tau},
\label{eq:appendix_A_5}
\end{eqnarray}
where $\rho_\chi$ is the density of dark matter, which is the same irrespective of the dark matter constituent, and $\tau$ is the lifetime of the particle constituting dark matter. The R.H.S. of the above equation has to be multiplied by the spectrum given off by each particle's decay to get the total injection rate of photons per unit volume. The $\tau$ and $m$ factors clearly cancel off. Thus, it does not matter what values of $\tau$ and $m$ we use, as long as they are reasonable values for particle dark matter.

Finally, the above discussion was done keeping in mind the injection of photons, for concreteness. But all the discussion can go through for $e^{\pm}$ also. We just have to replace the PPPC4DMID spectrum of $e^{\pm}$ also by our specified spectrum of them such that our imaginary decaying particle dark matter mimics the emission of $e^{\pm}$ by PBHs.

\subsection{Method B}
\label{sec: Method B}

This method relies on utilising \texttt{DarkHistory} and modifying particular functions within for PBH DM. Just like Method A, we take the emission spectra of evaporating PBHs from \texttt{BlackHawk} and use them as an inputs to \texttt{DarkHistory}.  Below we mention the changes that have to be made for PBH DM in files, \texttt{pppc.py} and \texttt{main.py}. Note that other \texttt{DarkHistory} files are left untouched. This is because only the energy injection has to be modified for the case of evaporating PBH DM, as can be seen from Eq.\,(\ref{eq:temp_ion_PBH}).

The default energy binning in \texttt{DarkHistory} is different than \texttt{BlackHawk}. So the spectra from \texttt{BlackHawk} have to be properly binned to the default energy binning of \texttt{DarkHistory}. Next, in the \texttt{pppc.py} file, for a particular PBH mass, we input the ($d^2N/dEdt$) in photon and electron-positron channels. Note that the default \texttt{Spectrum} function uses ($dN/dE$) instead of ($d^2N/dEdt$). Thus, we simply use the same $\Delta$t timestep specified in the \texttt{main.py} file of the default \texttt{DarkHistory} code, to convert $d^2N/dEdt$ to $dN/dE$. 

Next we modify the \texttt{main.py} file. For \texttt{main.evolve()} function, instead of the default particle DM inputs, we use our PBH model parameters, i.e., $M_{\rm PBH}$ and $f_{\rm PBH}$. We load the PBH spectra from the modified \texttt{pppc.py} using the functions \texttt{in\_spec\_elec} and \texttt{in\_spec\_phot}. The function \texttt{rate\_func\_N(rs)} calculates the rate of PBH evaporation per unit volume per unit time, given by,
\begin{eqnarray}
\left(\frac{dN}{dVdt}\right)=\frac{n_{\rm PBH}}{\Delta t}\,,
\label{eq:mainpy}
\end{eqnarray}
where, $n_{\rm PBH}$ is defined in Eq.\,(\ref{eq: number density PBH}). We replace the default formula for \texttt{rate\_func\_N(rs)} in \texttt{main.py} file with Eq.\,(\ref{eq:mainpy}). Similarly, the function \texttt{rate\_func\_eng(rs)} calculates the total energy injected per unit volume per unit time by PBH evaporation, 

\begin{align}
\left(\frac{dE}{dV dt}\right)_{\rm inj}^\gamma &= n_{\rm PBH} \left( \int E_\gamma \left(\frac{d^2 N}{dE dt}\right)_\gamma \, dE_\gamma \right. \nonumber\\
&\left. + \int 2(E_e - m_e c^2)\left(\frac{d^2 N}{dE dt}\right)_{e^\pm} dE_e \right) \,.
\end{align}
The integral of the Hawking evaporation spectra over energy is done using the default \texttt{DarkHistory} functions, \texttt{in\_spec\_elec.toteng()} and \texttt{in\_spec\_phot.toteng()}. Finally one has to remember that because this is Hawking evaporation, there will always be an electron-positron spectra. Thus one has to keep ‘\texttt{elec\_processes} = True’ in the \texttt{main.evolve()} function.

With these changes in place one can use the \texttt{main.evolve()} to obtain the temperature and ionization evolution for evaporating PBH DM. 
\vspace{10px}

Methods A (\ref{sec: Method A}) and B (\ref{sec: Method B}) are valid for both non-spinning and spinning PBHs. The only difference will be in the spectra obtained from \texttt{BlackHawk}. 

For any additional queries regarding the modifications of \texttt{DarkHistory} for PBH (using any of the two methods), we urge the reader to contact the authors.

\bibliographystyle{JHEP}
\bibliography{ref.bib}

\end{document}